\documentclass{article}
\usepackage[utf8]{inputenc}
\usepackage[affil-it]{authblk} % Package for affiliations
\usepackage{amsmath}
\usepackage{amsfonts} 
\usepackage{graphicx}
\usepackage{geometry}
\usepackage{hyperref}
\usepackage{subcaption}
\usepackage{lineno}
\usepackage{color}
\usepackage{booktabs}
\usepackage{url} 
\usepackage{setspace}
\usepackage{xcolor}
\usepackage{tabularx}
\usepackage{multirow}
\usepackage{natbib}
\usepackage{float}

\title{Copula and spatial-regularized variational autoencoder for mapping disease comorbidity in West Africa}
\author[1]{Osafu Augustine Egbon\thanks{Corresponding author: 111 T.W. Alexander Drive, Durham, 27709, USA; \texttt{eosafu.a@gmail.com}}}
\author[2]{Bassey David Ita}
\author[2]{Faith Eshofonie}
\author[2]{Ezra Gayawan}

\affil[1]{Institute of Mathematics and Computer Sciences, University of São Paulo, São Carlos, Brazil}
\affil[2]{The Federal University of Technology, Akure, Nigeria}

\begin{document}
\maketitle
\begin{abstract}
Geospatial health disproportionality remains a critical public health concern, as communities face heterogeneous illness risks due to varying exposures to adverse socioeconomic and environmental conditions.  While statistical models have been adopted to identify risk factors, studies that account for the complex, non-linear dependencies and spatial regularities inherent in comorbid disease patterns are underdeveloped. In this work, we propose a novel spatially regularized variational autoencoder (VAE) to characterize and map the geospatial disproportion of childhood comorbidity in West Africa, focusing on diarrhea, fever, and acute respiratory infection (ARI). To model dependence between these conditions, this study integrates a bivariate Gumbel copula into the VAE framework, enabling flexible modeling of asymmetric dependence and quantification of joint and conditional morbidity risks. Additionally, covariate effects within the framework were quantified to facilitate epidemiological interpretation of risk factors. The proposed method was benchmarked against commonly used methods and applied to characterize comorbidity in West Africa using the Demographic and Health Survey data. Findings reveal pronounced spatial heterogeneity in the likelihood of comorbidity among West African children, with the strongest co-occurrence observed between fever and ARI. Household wealth, maternal education, and access to improved water sources were associated with the likelihood of comorbidity. These patterns highlight high-risk areas and underscore the need for targeted, location-specific public health interventions.
\end{abstract}
\textbf{Keywords}: Deep learning, Geospatial modeling, Gumbel copula, Neural network, West Africa.
% \jnlcitation{\cname{%
% \author{Williams K.}, 
% \author{B. Hoskins}, 
% \author{R. Lee}, 
% \author{G. Masato}, and 
% \author{T. Woollings}} (\cyear{2016}), 
% \ctitle{A regime analysis of Atlantic winter jet variability applied to evaluate HadGEM3-GC2}, \cjournal{Q.J.R. Meteorol. Soc.}, \cvol{2017;00:1--6}.}

% \footnotetext{\textbf{Abbreviations:} ANA, anti-nuclear antibodies; APC, antigen-presenting cells; IRF, interferon regulatory factor}
\newpage
\section{Introduction}
Health disproportionality remains a significant public health challenge, especially in low- and middle-income countries such as those in West Africa, with illnesses such as fever, diarrhea, and acute respiratory infection (ARI), accounting for a significant burden among children under five \citep{demoze2025comorbidity,Gayawan2023Copula}. These illnesses are associated with the complex interaction of socioeconomic, demographic, and environmental exposures \citep{Adedokun2020,Orunmoluyi2022}. Despite decades of intervention, the geographic distribution of these health outcomes remains uneven, reflecting deep-seated structural inequities. These inequalities are more pronounced in comorbid health conditions.  Geospatial quantification of this disproportionality is essential for elucidating the likely structural factors contributing to this heterogeneity and guiding targeted interventions.

Comorbidity is the coexistence of two or more diseases or conditions in the same individual and is associated with worse health outcomes, more complex clinical management, and increased health care costs \citep{valderas2009defining}. Most comorbidity research has focused on chronic conditions in older adults with relatively limited attention to comorbidity patterns in children under five years of age. Using data from the Demographic and Health Survey across West African countries in this study, we observed substantial overlap between common childhood illnesses. Among children with fever, 31.2\% also had diarrhea, and among those with symptoms of ARI, 30.1\% had diarrhea. In addition, 53.6\% of children with ARI reported concurrent fever. We also observed that the co-occurrence substantially varies across communities, with a substantially higher proportion in regions with limited access to health facilities. These descriptive estimates highlight that, even in early childhood, comorbidity of acute conditions such as fever, diarrhea, and ARI is frequent and should be explicitly considered in epidemiologic analyses and in the design of integrated child health interventions in low- and middle-income settings.

Recent research has extensively applied statistical models to characterize the prevalence and determinants of diarrhea, acute respiratory infection (ARI), and fever, particularly in low- and middle-income countries \citep{hana_2024, cheng_2022, rahman2022prevalence, akter2025prevalence, fekadu2025trends, gayawaninvestigating}. However, approaches for formally quantifying comorbidity among these conditions remain relatively underdeveloped, despite growing recognition that disease co-occurrence patterns can reveal important shared mechanisms. Motivated by the need to infer and map the geospatial structure of comorbidity for diarrhea, ARI, and fever, this study integrates classical statistical methodology for disease co-occurrence with a deep learning framework, aiming to capture both local dependence and complex spatial patterns that are not easily accommodated by traditional models alone.

Specifically, this study proposes a spatially regularized deep learning framework based on variational autoencoders (VAEs) \citep{Simidjievski_2019,Hira2021} to integrate household survey data and geospatial covariates within a bivariate Gumbel copula framework to capture the pairwise dependencies between diarrhea, fever, and ARI among children in West African countries, and estimate the joint and conditional probabilities of these ailments. The Gumbel copula was adopted to capture asymmetric dependence, allowing weak association among absent symptoms while emphasizing clustering of co-morbidity \citep{nelsen2006introduction,joe2014dependence,lambert2002copula}. The Gaussian Markov Random Field (GMRF) model was introduced within the framework to allow spatially correlated information to be shared among neighboring locations. In addition, the epidemiological relevance of risk factors to the comorbidity of these ailments was quantified. The findings from this study are resourceful for guiding policymaking and sustainable intervention planning.

\section{Materials and Methods}
\subsection{Data}
We relied on DHS data from 14 Western African countries. The surveys are conducted at five years interval by the government agency responsible for the Population and Housing Census or a statistical agency of the country, following standard methodologies, procedures, and manuals, such that the data collected are comparable across locations. Usually, a two-stage sampling design is adopted for data collection, where enumeration areas are delineated from available census frames at the first stage, followed by the selection of households. All women of reproductive age (age 15-49 years) and their children under five years of age in the selected households are eligible for the interview. 
The following countries (and their survey year) were included in the study: Benin (2017-2018), Burkina Faso (2021), Cameroon (2018–2019), Cote d'Ivoire (2021), The Gambia (2019-2020), Ghana (2022), Guinea (2018), Liberia (2019-2020), Mali (2018), Niger (2012), Nigeria (2018); Senegal (2023),
Sierra Leone (2019), and Togo (2013-2014). Similar to previous studies, we considered the influence of time misalignment on the occurrence of the illnesses of interest to be minimal because of the short durations between the surveys \citep{Gayawan2023Copula,dunn2020spatially}. 

Three binary outcome variables were considered: ARI, diarrhea, and fever. The variables were measured based on a two-week recall in which the caregiver was asked if the child suffered from or had symptoms of the illnesses during the two weeks prior to the survey. The following covariates were considered: type of place of residence, educational attainment, source of drinking water (improved or unimproved), type of toilet facility, whether or not the household has electricity, where or the caregiver accesses each of radio, television and newspaper at least once in a week, household wealth index, whether or not the caregiver was working, the child's birth order, sex, age (in months) and whether or not the child slept under bed net in the night before the survey, and the caregiver's age (in years). The county, region, or
state where the child lived at the time of the survey was used as the spatial unit. There are 168 spatial units in the 14 Western African countries.

\subsection{Spatially informed VAE architecture}
Variational autoencoder provides a robust framework for modeling complex stochastic processes \citep{vincent2008extracting,makhzani2013k,kingma2014adam}, facilitating both predictive accuracy and formal statistical inference. While rooted in deep learning architectures, the VAE's probabilistic foundation enables seamless integration with classical statistical models. This hybrid nature affords a degree of interpretability and structural transparency often absent in purely frequentist machine learning methods.

In this work, we propose a spatially regularized variational autoencoder with a bivariate copula-based likelihood to jointly model correlated binary health outcomes (diarrhea, fever, and ARI) in multiple pairwise models while accounting for spatial dependencies across spatial units where the data were collected. The framework integrates representation learning, correlation modeling, and spatial smoothness within a single probabilistic deep learning architecture. %In the online supplementary material, we described the fundamentals of the variational autoencoder framework and the related works from a deep learning perspective for mapping different health outcomes.

\begin{figure}
    \centering
    \includegraphics[scale=0.49]{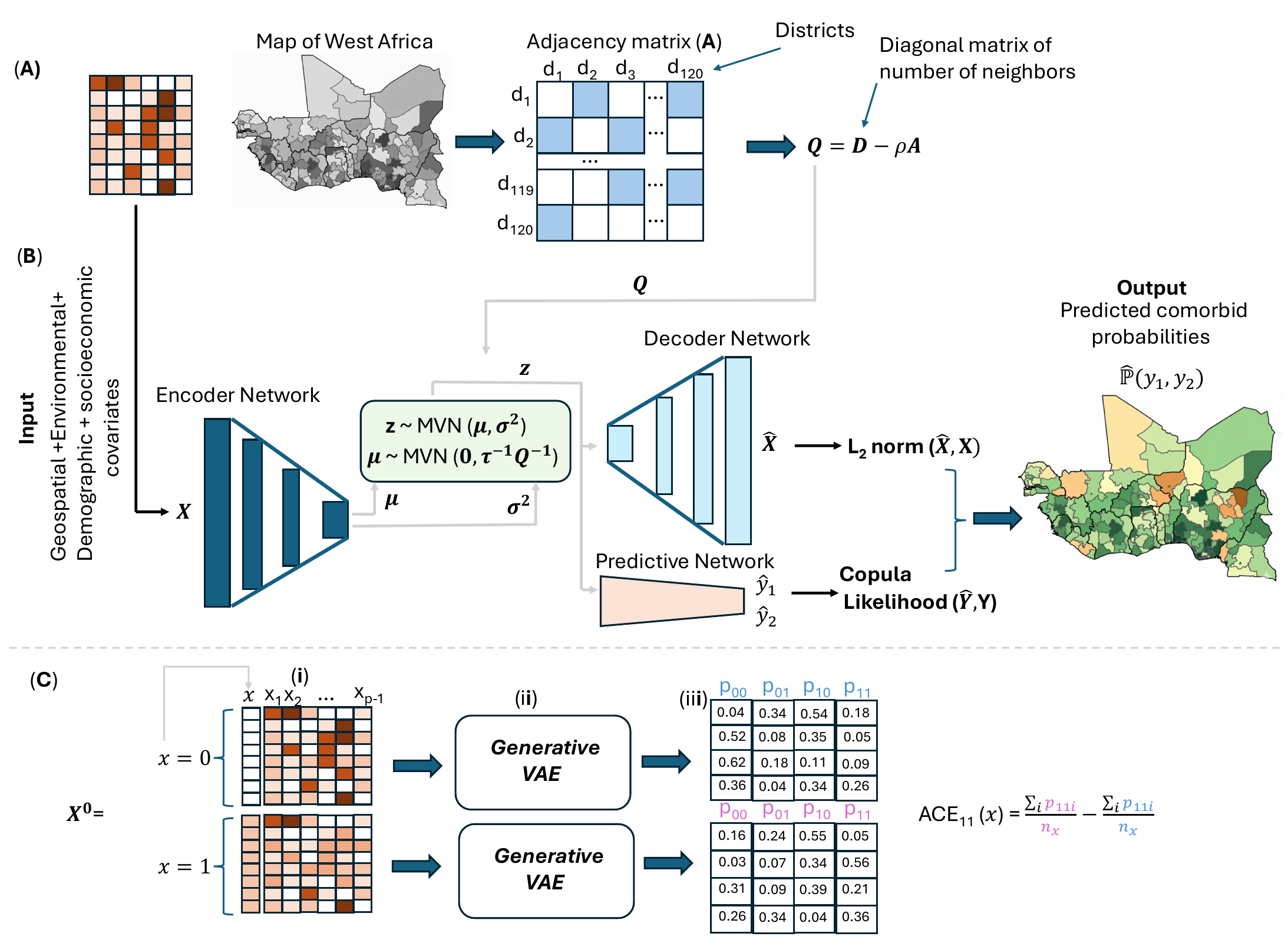}
    \caption{Schematic diagram of the proposed spatially regularized copula variational autoencoder (VAE) framework.
(a) The input data include geospatial, environmental, demographic, and socioeconomic covariates.
(a ii) The map of West Africa is represented as an adjacency matrix $\mathbf{A}$, which is used to derive the spatial precision matrix $\mathbf{Q}$ governing regional dependence.
(b) The encoder network comprises multiple hidden layers that learn latent representations $\boldsymbol{\mu}$ and $\boldsymbol{\sigma}^2$, where spatial smoothing of $\boldsymbol{\mu}$ is imposed through a Gaussian Markov random field prior derived from $\mathbf{Q}$. The schematic also depicts the decoding and predictive sub-networks that reconstruct inputs and jointly predict binary outcomes.
(c) The final panel illustrates the estimation of the average covariate effect (ACE) for a categorical covariate with two levels (0 and 1).
(c i–ii) The generative model is applied to subsets of the input data to generate predicted probabilities, which are then used to estimate the ACE and its bootstrap-based confidence intervals.}
    \label{fig:placeholder}
\end{figure}

Figure \ref{fig:placeholder} illustrates the overview of the architecture of the proposed spatially regularized VAE model.
Panel (a) depicts the input data, consisting of demographic, socioeconomic, environmental, and geospatial covariates, denoted by $\mathbf{X}$. The geospatial component is represented through a West Africa adjacency matrix $\mathbf{A}$, from which the spatial precision matrix $\mathbf{Q} = \mathbf{D} - \rho \mathbf{A}$ is derived. Here, $\mathbf{D}$ is a diagonal matrix whose entries correspond to the number of neighboring districts, and $\rho$ ensures that $\mathbf{Q}$ is positive definite \citep{besag1974spatial}. Panel (b) shows the encoder network, which consists of multiple hidden layers that learn a latent representation from the high-dimensional and heterogeneous input data. The encoder outputs two latent variables: the location (mean) parameter $\boldsymbol{\mu}$ and the scale (variance) parameter $\boldsymbol{\sigma}^2$. To incorporate spatial dependence, we impose a Gaussian Markov Random Field (GMRF) prior on $\boldsymbol{\mu}$, defined as $\boldsymbol{\mu} \sim \text{MVN}(\mathbf{0}, \tau^{-1}\mathbf{Q}^{-1})$, allowing information to be shared among neighboring districts and thereby enforcing spatial smoothness. The same panel also depicts the decoder and predictive networks, which transform the latent representation $\mathbf{z}$ back to the reconstructed input $\mathbf{X}$ and to the prediction of the bivariate target variable $\mathbf{Y}$ via a copula-based likelihood. The mapping back to $\mathbf X$ ensures that the latent representation captures the heterogeneity in $\mathbf X$ as much as possible for predicting the target outcomes. After model optimization, the learned latent variables $\mathbf{z}$ are used to generate probabilistic projections of comorbidities, including fever-ARI, fever-diarrhea, and ARI-diarrhea pairs. Panel (c) of the figure illustrates the estimation of the average covariate effect using the developed generative model for quantifying the epidemiological relevance of covariates. The theoretical details of the proposed architecture are presented in the next subsections. 

\subsubsection{Stochastic formulation}\label{latentsec}

Although the primary objective is the prediction of the joint probability of health outcomes given observed covariates (say \texttt{P(Y|X)}), we adopt a joint generative modeling framework in which both covariates and responses arise from a shared latent variable (say \texttt{P(Y, X|Z)}). This approach provides several advantages. First, it enables nonlinear denoising of high-dimensional covariates, leading to more stable and interpretable latent representations \citep{kingma2013auto}. Second, the inclusion of a covariate likelihood acts as a form of regularization, preventing overfitting to the response and improving generalization \citep{murphy2012machine}. Third, the joint modeling naturally accommodates measurement error and latent heterogeneity in covariates, which is common in spatial and environmental health data. Finally, modeling the joint distribution allows flexible dependence structures to be integrated coherently within a unified probabilistic model. 

 Let  $\mathbf x_{ij}\in \mathbb R^p$ represent the covariate vector and $\mathbf y_ {ij}=(y_{ij1},y_{ij2})^\top$ represent the corresponding health condition, where $i (i=1,2,...,n_j)$ represents the observation at district $j (j=1,2,...,L)$. Let  $\mathbf z_{ij}\in \mathbb {R}^d$ be a $d$-dimensional latent variable such that $\mathbf z_{ij}= (z_{ij1},z_{ij2},...,z_{ijd})$. We assume that the covariate vector $\mathbf x_{ij}$ and target response vector $\mathbf y_{ij}$ are independent conditional on $\mathbf z_{ij}$ and were generated from a distribution conditioned on $\mathbf z_{ij}$ through the following stochastic model

\begin{align}
\begin{aligned}
   \boldsymbol\mu_{k}= (\mu_{1k},\mu_{2k},...,\mu_{Lk})^\top &\sim N(\mathbf 0, \tau^{-1}\mathbf Q^{-1}),\, \mu_{jk}\in \mathbb R, k=1,2,...,d,\\
    z_{ijk} &\sim N(\mu_{jk}, \sigma^2_{jk}),\,\\
    \mathbf x_{ij}\mid  \mathbf z_{ij} &\sim p_{\theta_x}(\mathbf x_{ij}\mid  \mathbf z_{ij})\\
      (y_{ij1},y_{ij2})^\top\mid \mathbf z_{ij}&\sim p_{\theta_y}(y_{ij1},y_{ij2}\mid\mathbf z_{ij}),
      \end{aligned}
      \label{hierarchical}
\end{align}
where $p_{\theta_x}$ and $p_{\theta_y}$ are probability distributions of the observed data, $\mathbf Q$ is the spatial Laplacian precision matrix as previously defined, with global scaling parameter $\tau$. Specific for this application, we assumed $p_{\theta_x}$ is a Gaussian distribution and $p_{\theta_y}$ is a joint distribution constructed via a copula distribution for $y_{ij1}$ and $y_{ij2}$. In practice, each component of $\mathbf x_{ij}$ was normalized across $ij$ observations. The copula model incorporates the dependency between the response variables.

The conditional independence assumption $\mathbf x_{ij} \perp \mathbf y_{ij} \mid \mathbf z_{ij}$ concentrates all shared information into a low-dimensional latent representation, $\mathbf z_{ij}$, simplifying joint modeling and improving statistical efficiency. In addition, the spatial prior $\boldsymbol\mu_k \sim \mathcal N(\mathbf 0, \tau^{-1}\mathbf Q^{-1})$ enforces smoothness across neighboring districts, enabling borrowing of information and stabilizing estimates in sparse regions. The hierarchical latent layer $z_{ijk} \sim \mathcal N(\mu_{jk}, \sigma^2_{jk})$ decomposes variability into district-level structure and individual-level noise, thereby capturing multilevel heterogeneity. The covariate model $p_{\theta_x}(\mathbf x_{ij}\mid \mathbf z_{ij})$ regularizes the latent space by forcing it to explain observed exposures, improving robustness to noise and high dimensionality. The response model $p_{\theta_y}(y_{ij1},y_{ij2}\mid \mathbf z_{ij})$ enables flexible prediction of outcomes from latent factors while leveraging shared structure with covariates. Moreover, the copula construction for $(y_{ij1},y_{ij2})$ explicitly captures dependence between outcomes, improving joint risk estimation and modeling of comorbidity.

The full joint distribution can be expressed as
%SEEE CHAT https://chatgpt.com/share/69d152ee-dd7c-8332-ab70-cd95dd008477}
\begin{align}
    p(\mathbf x, \mathbf y, \mathbf z, \boldsymbol\mu)=
\prod_{k=1}^d p(\boldsymbol\mu_{k})
\prod_{j=1}^L \prod_{i=1}^{n_j}
p_{\theta_x}(\mathbf x_{ij} \mid \mathbf z_{ij})
p_{\theta_y}(\mathbf y_{ij} \mid \mathbf z_{ij})\prod_{k=1}^d\prod_{j=1}^L \prod_{i=1}^{n_j}p(z_{ijk} \mid \mu_{jk}).
\label{jng}
\end{align}

However, we wish to maximize the joint marginal distribution of the observed data given as
\begin{align}
   p(\mathbf x, \mathbf y)=
 \int p(\mathbf x, \mathbf y, \mathbf z, \boldsymbol\mu) d\mathbf z d\boldsymbol\mu.
\label{jointmarginal}
\end{align}
$p(\mathbf x, \mathbf y)$ is complex and intractable due to the copula model. Hence, we adopt a variational approximation method through a variational autoencoder framework. %described in the next subsections.
%where $\Phi_2(\cdot,\cdot;\rho)$ denotes the bivariate standard normal CDF with correlation $\rho$, and $\tilde{\eta}_{ij} = (2y_{ij} - 1)\eta_{ij}$.
\subsection{Variational inference}
To begin, we assumed a variational joint posterior distribution for $\mathbf z$ and $\boldsymbol\mu$ given as
\begin{align}
q(\mathbf z, \boldsymbol\mu \mid \mathbf x)
=q(\boldsymbol\mu)q(\mathbf z \mid \mathbf x).%=\prod_{k=1}^dq(\boldsymbol\mu_k)\prod_{k=1}^d\prod_{i=1}^{n_j}q_\phi(\mathbf z_{ik} \mid \mathbf x).
\label{encoder0}
\end{align}

For computational tractability, we adopt a posterior point estimate for the spatial latent field by specifying
\begin{align}
q(\boldsymbol\mu) = \delta(\boldsymbol\mu - \hat{\boldsymbol\mu}),
\end{align}
where $\delta$ is the Dirac function,  which places a mass at $\boldsymbol\mu$ and zero everywhere. $\hat{\boldsymbol\mu} = \mathbb E_{q(\boldsymbol\mu)}(\boldsymbol\mu)$. %which reduces the corresponding Kullback–Leibler divergence to the negative log-prior.
%We assumed a Dirac delta function for $q(\boldsymbol\mu)=\delta(\boldsymbol\mu)$, which places a mass at $\boldsymbol\mu$ and zero everywhere. 
In addition, we assumed a conditional distribution for $\mathbf z$ given as
\begin{align}
    q(\mathbf z_{ij} \mid \mathbf x) =\mathcal N(\boldsymbol\beta_{ij}, \mathrm{diag}(\boldsymbol\sigma_{ij}^2)),
    \label{encoder}
\end{align}
where $\boldsymbol\beta_{ij}$ and $\boldsymbol\sigma_{ij}^2$ are functions of $\mathbf x$ derived from an encoder neural network using the reparameterization trick (see section \ref{latentsec}). Applying Jensen's inequality for concave functions, the joint distribution can be rewritten as

\begin{align}
\begin{aligned}
     p(\mathbf x, \mathbf y)&=
\int  p(\mathbf x, \mathbf y, \mathbf z, \boldsymbol\mu) \frac{q(\mathbf z, \boldsymbol\mu \mid \mathbf x)}{q(\mathbf z, \boldsymbol\mu \mid \mathbf x)}d\mathbf z d\boldsymbol\mu= \mathbb E_{q(\mathbf z,\boldsymbol\mu\mid \mathbf x)}\left[\frac{ p(\mathbf x, \mathbf y, \mathbf z, \boldsymbol\mu)}{q(\mathbf z, \boldsymbol\mu \mid \mathbf x)} \right], \text{ and }\\
 \log p(\mathbf x, \mathbf y)&\geq  \mathbb E_{q(\mathbf z,\boldsymbol\mu\mid \mathbf x)}\left[\log \frac{ p(\mathbf x, \mathbf y, \mathbf z, \boldsymbol\mu)}{q(\mathbf z, \boldsymbol\mu \mid \mathbf x)} \right].
 \end{aligned}
\label{jointmarginal2}
\end{align}
Expanding the joint distribution as given in \eqref{jng}, we have 
\begin{align}
\begin{aligned}
   \log p(\mathbf x, \mathbf y)\geq& \mathbb E_{q(\mathbf z,\boldsymbol\mu\mid \mathbf x)} \left[\log p(\mathbf x\mid \mathbf z)\right]+\mathbb E_{q(\mathbf z,\boldsymbol\mu\mid \mathbf x)} \left[\log p(\mathbf y\mid \mathbf z)\right]+\\
   & \mathbb E_{q(\mathbf z,\boldsymbol\mu\mid \mathbf x)} \left[\log p(\mathbf z\mid \boldsymbol\mu)\right] +  \mathbb E_{q(\mathbf z,\boldsymbol\mu\mid \mathbf x)} \left[\log p( \boldsymbol\mu)\right]- \mathbb E_{q(\mathbf z,\boldsymbol\mu\mid \mathbf x)}\log q(\mathbf z,\boldsymbol\mu\mid \mathbf x).
   \end{aligned}
   \label{h2}
\end{align}
Since $\mathbb E_{q(\mathbf z,\boldsymbol\mu\mid \mathbf x)}\log q(\mathbf z,\boldsymbol\mu\mid \mathbf x) = \mathbb E_{q(\mathbf z\mid \mathbf x)}\log q(\mathbf z\mid \mathbf x)+ \mathbb E_{q(\boldsymbol\mu)}\log q(\boldsymbol\mu)$. Grouping \eqref{h2}, we have
\begin{align}
\begin{aligned}
   \log p(\mathbf x, \mathbf y)\geq& \mathbb E_{q(\mathbf z\mid \mathbf x)} \left[\log p(\mathbf x\mid \mathbf z)\right]+\mathbb E_{q(\mathbf z\mid \mathbf x)} \left[\log p(\mathbf y\mid \mathbf z)\right]-\\
   & \mathbb E_{q(\mathbf z\mid \mathbf x)} \left[\log q(\mathbf z\mid \mathbf x) -\log p(\mathbf z\mid \boldsymbol\mu)\right]-  \mathbb E_{q(\boldsymbol\mu)} \left[ \log q(\boldsymbol\mu) -\log p( \boldsymbol\mu)\right].
   \end{aligned}
   \label{h22}
\end{align}
Finally, 

\begin{align}
\begin{aligned}
     \log p(\mathbf x, \mathbf y)\geq&  
     \mathbb E_{q(\mathbf z\mid \mathbf x)} \left[\log p(\mathbf x\mid \mathbf z)\right]+\mathbb E_{q(\mathbf z\mid \mathbf x)} \left[\log p(\mathbf y\mid \mathbf z)\right]-\\
     &\text{KL}\left[q(\mathbf z\mid \mathbf x)\mid\mid p(\mathbf z\mid \boldsymbol\mu) \right ]-\text{KL}\left[q(\boldsymbol\mu)||p(\boldsymbol\mu)\right ]=\text{ELBO},
     \end{aligned}
     \label{elbo}
\end{align}
where $\boldsymbol\mu$ is evaluated at the point mass $\hat{\boldsymbol\mu}$ and ELBO is the Evidence Lower Bound. The KL$(q||p)$ represents the Kullback-Leibler divergence, or relative entropy, which measures how an approximate probability distribution $q$ differs from a true probability distribution $p$. It measures the expected excess surprise or information loss when using $q$ to approximate $p$. Details of the derivation of the full KL divergence are shown in the Supplementary material.  Maximizing the log-likelihood $ \log p(\mathbf x, \mathbf y)$ with respect to the model parameters is equivalent to maximizing the ELBO on the right-hand side of \eqref{elbo}. 

%We leveraged neural network model known as the encoder and a reparameterization trick to obtain $ q_\phi(\mathbf z_{ij}\mid \mathbf x_{ij})$ in Equation \eqref{encoder}. Similarly, we used decoder neural network to derive $p_{\theta_x}$, and a fed-forward neural network to derive $p_{\theta_y}$. The details of the deep neural network are presented briefly in the subsequent section.

Recall from Equation \eqref{hierarchical} that 
 $\mathbf x_{ij}\mid  \mathbf z_{ij} \sim p_{\theta_x}(\mathbf x_{ij}\mid  \mathbf z_{ij})$ and $(y_{ij1},y_{ij2})^\top\sim p_{\theta_y}(y_{ij1},y_{ij2}\mid\mathbf z_{ij})$, and from Equation \eqref{encoder0} that the joint distribution for all samples $q(\mathbf z, \boldsymbol\mu \mid \mathbf x)
=q(\boldsymbol\mu) q(\mathbf z \mid \mathbf x)$. We employed a neural network to approximate the variational posterior distribution for $q(\mathbf z_{ij} \mid \mathbf x_{ij})$, known as the encoder. The encoder takes an input $\mathbf{x}_{ij}$ and learns a mapping to the variational posterior distribution $q(\mathbf{z}_{ij}|\mathbf{x}_{ij})$. The encoder's primary job is to find the most essential features of the data. It forces the input through a bottleneck and spatial smoother (the latent space $\mathbf{z}_{ij}$), and must learn to discard noise and retain only the variables necessary. The spatial prior $\boldsymbol{\mu}$ further regularizes this bottleneck by ensuring that the latent representations of neighboring districts are consistent, effectively denoising the signal across both the feature space and the geospatial domain.

The generative component, $p_{\theta_x}(\mathbf{x}_{ij} \mid \mathbf{z}_{ij})$, serves as the decoder. It maps the compressed latent representation $\mathbf{z}_{ij}$ back into the high-dimensional observation space to reconstruct the original data $\mathbf{x}_{ij}$. During training, the optimization problem enforces an information bottleneck that necessitates $\mathbf{z}_{ij}$ to preserve the salient structural patterns of the dataset while effectively filtering out stochastic noise. The predictive component, $(y_{ij1}, y_{ij2})^\top \sim p_{\theta_y}(y_{ij1}, y_{ij2} \mid \mathbf{z}_{ij})$, functions as the prediction network, where the joint dependency between outcomes is modeled via a copula-based distribution. This framework allows the network to capture complex, non-linear correlations between the response variables conditioned on the latent manifold $\mathbf{z}_{ij}$. In the next subsection, we detail the encoder, decoder, and predictive components.
\subsubsection{Encoder and decoder}\label{latentsec}

Let $\mathbf{x}_{ij} \in \mathbb{R}^p$ denotes the vector of predictors for observation $i$, and $\mathbf{y}_{ij} = (y_{ij1}, y_{ij2})^\top$ be the binary outcomes, say, fever $(y_{ij1})$ and diarrhea $(y_{ij2})$, where $y_{ij1}=1$ indicates that child $i$ in location $j$ has the phenotype fever and $y_{ij2}=1$ denotes that the same child has diarrhea. We assume that $\mathbf{x}_{ij}$ can be represented by a lower-dimensional latent variable $\mathbf{z}_{ij} \in \mathbb{R}^d,d<p$ through a probabilistic mapping
\begin{equation}
    \mathbf{z}_{ij}\mid \mathbf x_{ij} \sim \mathcal{N}(\boldsymbol{\beta}_{ij}, \operatorname{diag}(\boldsymbol{\sigma}_{ij}^2)),
\end{equation}
where $\boldsymbol{\beta}_{ij}$ and $\boldsymbol{\sigma}_{ij}$ are functions of $ \mathbf{h}_{ij}^{(e_{I})}$ derived from an encoder neural network $ \mathbf{h}_{ij}^{(e_{I})} = f_{\text{enc}}(\mathbf x_{ij};\boldsymbol\theta_e)$ with $I$ hidden layers that maps $\mathbf{x}_{ij}$ to its latent vector. The encoder $f_{\text{enc}}$ is defined sequentially as
\begin{align}
\begin{aligned}
    \mathbf{h}_{ij}^{(e_{t+1})} &= \text{ReLU}\left(\mathbf{W}^{(e_{t})}\mathbf{h}_{ij}^{({e_t})} + \mathbf{b}^{(e_{t})}\right),\, \mathbf h^{(e_t)}_{ij}\in\mathbb R^{d_t},\\
   t&=0,1,2,...,I-1,\\
    \mathbf h^{(e_0)}_{ij}&=\mathbf x_{ij} \text{ and } \dim(\mathbf{h}^{(e_{t+1})})<\dim(\mathbf{h}^{(e_{t})}),
        \end{aligned}
\end{align}
where $\text{ReLU}(a) = \max(0, a)$ is an activation function, which makes neural networks efficient at learning complex patterns through allowing passage of positive signals while suppressing negative ones, speeding up training, and helping avoid the vanishing gradient problem. The matrix $\mathbf{W}^{(e_t)}\in \mathbb R^{d_{t+1}\times d_{t}}$ is the network weights and $\mathbf b^{(e_t)}\in \mathbb R^{d_{t+1}}$ is the vector of the bias. For a specific low-dimensional representation at $t=I$, we derive the latent location parameter $\boldsymbol\mu_{ij}$ and the scale parameter $\boldsymbol\kappa_{ij}$ in the log scale given as \citep{dilokthanakul2016deep}.
\begin{align}
\begin{aligned}
    \boldsymbol{\beta}_{ij} &= \mathbf{W}^{(\mu)} \mathbf{h}_{ij}^{({e_{I}})} + \mathbf{b}^{(\mu)},\\
    \boldsymbol{\kappa}_{ij} &= \mathbf{W}^{(\sigma)} \mathbf{h}_{ij}^{(e_{I})} + \mathbf{b}^{(\sigma)},\,\text{where}\\
    \boldsymbol{\beta}_{ij},\boldsymbol{\kappa}_{ij}&\in \mathbb R^{d}.
    \end{aligned}
    \label{enc}
\end{align}
%where $\mathbf h^{(0)}_i=\mathbf x_i$. ReLU$(a)$ is a function that returns $a$ if $a>0$ and $0$ other wise, 
Equation \ref{enc} is similar in spirit to a distributional regression in statistical modeling, where the mean $ \boldsymbol{\beta}_{ij}$ and variance $\boldsymbol{\sigma}_i^2=\exp(\boldsymbol\kappa_{ij})$ are both linked to covariates through two different predictors. The neural network is different from the statistical regression equivalent mostly due to the non-linear and complex links of the covariates to the response variables. 

In practice, the latent representation output, $\mathbf{z}_i$, is obtained from a reparameterization trick, given as 
\begin{equation}
    \mathbf{z}_{ij} = \boldsymbol{\beta}_{ij} + \exp(0.5 \, \boldsymbol{\kappa}_{ij}) \odot\boldsymbol{\epsilon}_{ij},
    \quad \boldsymbol{\epsilon}_{ij} \sim \mathcal{N}(\mathbf{0}, \mathbf{I}),
\end{equation}
ensuring differentiability with respect to $\boldsymbol\theta_e=\{\mathbf W,\mathbf b\}$. This reparameterization trick speeds up the stochastic gradient optimization routine to obtain $\hat {\boldsymbol\theta}_e$.

In addition, we define the decoder network to reconstruct $\mathbf x_{ij}$ from $\mathbf z_{ij}$ as $\hat{\mathbf x}_{ij}=f_{\text{dec}}(\mathbf z_{ij};\boldsymbol\theta_d)$ with $I$ hidden layers. We assumed a multivariate Gaussian distribution for $\mathbf x_{ij}$ with mean give as $\hat{\mathbf x}_{ij}=\mathbb E(\mathbf x_{ij}\mid \mathbf z_{ij})$. The decoder network $f_{\text{dec}}$ is defined sequentially as
\begin{align}
\begin{aligned}
    \mathbf{h}_{ij}^{(g_{t+1})} &= \text{ReLU}\left(\mathbf{W}^{(g_{t})}\mathbf{h}_{ij}^{(g_{t})} + \mathbf{b}^{(g_{t})}\right),\,   \mathbf h^{(g_t)}_{ij}\in\mathbb R^{d_t'},\\
    t&=0,1,2,...,I-1,\, \text{ and } \dim{(\mathbf h^{(g_t)}_{ij})}<\dim{(\mathbf h^{(g_{t+1})}_{ij})}, \,\mathbf h_{ij}^{(g_0)} = \mathbf z_i,\\
    \hat{\mathbf{x}}_{ij} &= \mathbf{W}^{(g_{I})} \mathbf{h}_{ij}^{(g_{I})} + \mathbf{b}^{(g_{I})},\, \\
   % \log \boldsymbol{\sigma}_i^2 &= \mathbf{W}^{(\sigma)} \mathbf{h}_i^{(e_{t'})} + \mathbf{b}^{(\sigma)}.
    \end{aligned}
    \label{dec}
\end{align}
where $\boldsymbol\theta_d =\{\mathbf W, \mathbf b\}$ for the decoder.

% The reconstruction loss encourages the prediction $\hat{\mathbf{x}}_i$ to match the input  ${\mathbf{x}}_i$. That is we seek to find $(\boldsymbol\theta_e,\boldsymbol\theta_d)^\top$ that minimixes
% \begin{equation}
%     \mathcal{L}_{\text{recon}} = \sum_{i=1}^{n} \| \mathbf{x}_i - \hat{\mathbf{x}}_i \|_2^2.
% \end{equation}
% Minimizing $\mathcal{L}_{\text{recon}}$ ensures that $\mathbf z_i$ mirrors the true latent representation of the high-dimensional covariates $\mathbf x_i$, resourceful for projecting the target variables $(y_{1i}, y_{2i})^\top$, presented in subsection \ref{seccopula}.

\subsubsection{Prediction network for bi-variate binary outcomes through Gumbel copula model}\label{seccopula}
From the latent representation $\mathbf{z}_{ij}$, we jointly predict the continuous latent variables $(\eta_{ij1}, \eta_{ij2})^\top\in \mathbb R^2$ underlying the binary responses from a feedforward neural network $(\eta_{ij1}, \eta_{ij2}) = f_{pred}(\mathbf z_{ij},\boldsymbol\theta_p)$. The sequential representation of $f_{pred}$ is given as
\begin{align}
\begin{aligned}
    \mathbf{h}_{ij}^{(p_{t+1})} &= \text{ReLU}\left(\mathbf{W}^{(p_{t})}\mathbf{h}_{ij}^{(p_{t})} + \mathbf{b}^{(p_{t})}\right),\, \mathbf h^{(p_t)}_{ij}\in\mathbb R^{d_t''} \\
    t&=0,1,2,...,I_L-1  \text{ and } \dim{(\mathbf h^{(p_{t+1})}_{ij})}<\dim{(\mathbf h^{(p_{t})}_{ij})},\\
   (\eta_{ij1},\eta_{ij2})^\top &= \mathbf{W}^{(p_{I_L})} \mathbf{h}_{ij}^{(p_{I_L})} + \mathbf{b}^{(p_{I_L})}.\,\\
    \end{aligned}
    \label{pred}
\end{align}

Since the copula model assumes its marginals are continuous and we want to leverage it to capture the dependence between the two binary outcomes, we adopted a latent variable model, commonly referred to as the probit model \citep{muthen1979structural}.  We define $Y^\star_{ij1} = \eta_{ij1} + \varepsilon_{ij1}$ corresponding to observed indicator  $y_{ij1}$ and $Y^\star_{ij2} = \eta_{ij2} + \varepsilon_{ij2}$ corresponding to observed indicator  $y_{ij2}$, where $\varepsilon_{ij1}$ and $\varepsilon_{ij2}$ are errors with standard normal distribution \citep{Gayawan2023Copula,briseno2025boosting}. We relate the following to the response as follows 

\begin{align}
Y_{ijl} = 
    \begin{cases}
       1 &\, \text{if }Y^\star_{ijl}>0,\\
        0&\,\text{if }Y^\star_{ijl}\leq0,\,l=1,2,
    \end{cases}
\end{align}
such that
\begin{align}
\begin{aligned}
P(Y_{ijl} = 1 \mid \theta_y) &= P(Y_{ijl}^\star > 0\mid \theta_y) \\
&= P(\eta_{ijl} + \varepsilon_{ijl} > 0\mid \theta_y) \\
%&= P(\varepsilon > - \eta_{1i}) \\
%&= P(\varepsilon < \eta_{1i}) \\
&= \Phi(\eta_{ijl}) = p_{ijl},
\end{aligned}
\end{align}
%where $Y^\star_{1} = \eta_{1i} + \varepsilon$ is a  latent variable \citep{gayawan2023copula,briseno2025boosting}. 
where $\Phi$ is the cumulative distribution of the standard normal distribution. By using $\Phi(\eta_{ijl})$, we effectively employ a Probit link function, which is naturally compatible with the Gaussian latent space of the VAE. However, we seek the joint probability distribution $P(Y_{ij1}=1, Y_{ij2}=1)$. Thus, we adopted a copula model, such that the joint probability is given as
\begin{align}
    P(Y_{ij1}=1, Y_{ij2}=1) = C(p_{ij1},p_{ij2}; \theta_y). %= \Phi_2(\Phi^{-1}(u_1),\Phi^{-1}(u_2);r),
\end{align}
%where $r$ is the correlation parameter. 

We adopted the Gumbel copula to model the pairwise dependence between diarrhea, ARI, and fever because these conditions are expected to exhibit strong upper-tail dependence, reflecting simultaneous escalation to severe disease states. The Gumbel copula captures asymmetric dependence, allowing weak association among absent symptoms while emphasizing clustering of extreme morbidity \citep{nelsen2006introduction,joe2014dependence,lambert2002copula}. The Gumbel copula distribution is given as
\begin{align}
    C_{\alpha}(p_1,p_2) = 
\exp\left(
-\left[
(-\log p_1)^{\alpha}
+
(-\log p_2)^{\alpha}
\right]^{1/\alpha}
\right),
\qquad
\alpha \ge 1,
\end{align}
with upper tail dependence given as $\lambda= 2 - 2^{1/\alpha}$. Thus,
%$\Phi_2$ and $\Phi$ are the bi-variate and uni-variate CDF of a standard Gaussian distribution, respectively. Thus, the joint probability 
\begin{align}
\begin{aligned}
        P(Y_{ij1} = 1 ,Y_{ij2} = 1) &=  C_{\theta_y}(p_{ij1},p_{ij2})=p_{ij11}.%\Phi_2(\Phi^{-1}( \Phi(\eta_{1i}) ),\Phi^{-1}(\Phi(\eta_{2i}));r)\\
%  &= \Phi_2(\eta_{1i},\eta_{2i};r) =.
\end{aligned}
\end{align}
To obtain $p_{ij10} := P(Y_{ij1} = 1 ,Y_{ij2} = 0)  =  p_{ij1} - p_{ij11}$ and $p_{ij01} := P(Y_{ij1} = 0 ,Y_{ij2} = 1)  =  p_{ij2} - p_{11i}$ and $p_{ij00} = 1-(p_{ij11}+p_{ij10}+p_{ij01})$. 

The joint distribution $p_{\theta_y}$ is then given as
\begin{equation}
   p_{\theta_y}(y_{ij1}, y_{ij2} \mid \mathbf{z}_{ij}) =  P(Y_{ij1} =  y_{ij1}, Y_{ij2}= y_{ij2})
    = p_{ij11}^{y_{ij1}y_{ij2}}p_{ij10}^{y_{ij1}(1-y_{ij2})}p_{ij01}^{(1-y_{ij1})y_{ij2}}p_{ij00}^{(1-y_{ij1})(1-y_{ij2})}.
\end{equation}

We adopted a regularization parameter $\lambda$ to weigh the amount of information from the reconstruction used in the prediction of the target variable. Thus, the ELBO is given as 

\begin{align}
\begin{aligned}  
     \text{ELBO} = E_{q(\mathbf z\mid \mathbf x)} \left[\log p(\mathbf y\mid \mathbf z)\right]+&\lambda\mathbb E_{q(\mathbf z\mid \mathbf x)} \left[\log p(\mathbf x\mid \mathbf z)\right]-\\
     &\text{KL}\left[q(\mathbf z\mid \mathbf x)\mid\mid p(\mathbf z\mid \boldsymbol\mu) \right ]-\text{KL}\left[q(\boldsymbol\mu)||p(\boldsymbol\mu)\right ],
     \end{aligned}
     \label{elbo2}
\end{align}

\noindent where $\mathbb{E}_{q(\mathbf{z} \mid \mathbf{x})} [\log p(\mathbf{y} \mid \mathbf{z})] = \sum_{j=1}^L \sum_{i=1}^{n_j} \mathbb{E}_{q(\mathbf{z}_{ij} \mid \mathbf{x}_{ij})} [\log p(\mathbf{y}_{ij} \mid \mathbf{z}_{ij})]$ and  $\mathbb{E}_{q(\mathbf{z} \mid \mathbf{x})} [\log p(\mathbf{x} \mid \mathbf{z})] = \sum_{j=1}^L \sum_{i=1}^{n_j} \mathbb{E}_{q(\mathbf{z}_{ij} \mid \mathbf{x}_{ij})} [\log p(\mathbf{x}_{ij} \mid \mathbf{z}_{ij})]$. $\lambda\in [0,1]$ acts as a hyperparameter for gradient balancing. Without it, the reconstruction likelihood might dominate the total likelihood, causing the model to ignore the labels $\mathbf{y}$ during the early stages of training. This is particularly important when there is a need for early stopping of the optimization. 

% The total loss combines the reconstruction, copula, spatial, and KL regularization terms:
% \begin{equation}
% \begin{aligned}
%     \mathcal{L}_{\text{total}}
%     &= \lambda_1\mathcal{L}_{\text{recon}}
%      + \lambda_2\mathcal{L}_{\text{copula}}
%      + \lambda_3\mathcal{L}_{\text{spatial}}
%      + \beta \, \mathcal{L}_{\text{KL}}, \\
%     \mathcal{L}_{\text{KL}}
%     &= -\frac{1}{2}\sum_{i=1}^{n}\sum_{k=1}^{d}
%     \left[ 1 +  {\kappa}_{ik} - {\mu}_{ik}^2 - \exp(\boldsymbol{\kappa}_{ik}) \right],
% \end{aligned}
% \end{equation}
% where $\lambda_1,\lambda_2, \lambda_2,\beta>0$ control the strength of regularization.
Model parameters 
$\Theta = \{\boldsymbol\theta_e, \boldsymbol\theta_d, \boldsymbol\theta_p, \alpha, \tau,\lambda\}$
are estimated by maximizing ELBO using the Adam optimizer with stochastic gradient descent \citep{reyad2023modified}. The estimation was performed using the \texttt{torch} R interface \citep{keydana2023deep}. $\alpha, \tau,$ and $\lambda$ are learned in the optimization process. Once optimized, the model is used to make posterior predictions and compute the posterior probabilities of the ailments across geospatial locations through the copula model.

%This model learns a spatially coherent latent representation of children's covariates that explains variation in two correlated disease outcomes through a shared latent space. The copula captures dependence between binary responses, while the spatial penalty encourages geographically smooth latent structures, thereby improving inference and predictive performance in spatially structured epidemiological data.

\subsection{Average covariate effects}
To quantify the epidemiological impact of each risk factor on the likelihood of a child experiencing comorbid diseases, we leverage the posterior predicted probabilities. For simplicity, suppose the covariate of interest is $x_{l} \in \mathbb R$ for component $l$ in  $\mathbf x$ in dataset $\mathcal D=\{\mathbf y_1,\mathbf y_2,\mathbf x\}$. We compute the covariate effect on \((Y_1, Y_2)\) as the difference in their joint probabilities under different exposure levels. We aim to test the hypothesis 
\begin{align}
    \mathcal H_0 : ACE_{y_1y_2}^{(x_l)} =0\,\, \text{vs}\,\,  \mathcal H_0 : ACE_{y_1y_2}^{(x_l)} \neq 0,
\end{align}
where
% For simplicity, suppose the covariate of interest is binary $x_{l} \in \{0,1\}$ for covariate $l$. We compute the covariate effect on \((Y_1, Y_2)\) as the difference in their joint probabilities under the two exposure levels 
\begin{align}
    ACE_{y_1y_2}^{(x_l)} 
    &= \mathbb{P}(Y_{1}=y_1,\,Y_{2}=y_2 \mid X_l=x_l,\mathcal D)
     -\mathbb{P}(Y_{1}=y_1,\,Y_{2}=y_2 \mid X_l=x_{ref},\mathcal D),
    \label{eq:causal_effect}
\end{align}
and \(\mathbf{x}^{-l}\) denotes the covariate vector but excluding the $l$th component corresponding to covariate $X_l$. $ACE_{y_1y_2}^{(x_l)}$ is the average change in predicted joint outcome probability when $X_l$ is shifted from $x_{ref}$ to $x_l$, and it is similar in spirit to Average Treatment Effect (ATE) \citep{schafer2008average,egbon2025malaria}. The posterior predictive probability $p^{y_1y_2(x_l)}= \mathbb{P}(Y_{1}=y_1,\,Y_{2}=y_2 \mid X_l=x_l,\mathcal D)$ is obtained through the integral

\begin{align}
\begin{aligned}
\mathbb{P}(Y_{1}=&y_1,\,Y_{2}=y_2 \mid X_l=x_l,\mathcal D) =\int \int\int \mathbb{P}(Y_{1}=y_1,\,Y_{2}=y_2, \mathbf X^{-l},\mathbf z,\boldsymbol\mu \mid X_l=x_l,\mathcal D) d p(\mathbf X^{-l})dp(\mathbf z)dp(\boldsymbol\mu)\\
=&\int \int\int \mathbb{P}(Y_{1}=y_1,\,Y_{2}=y_2\mid \mathbf z,\boldsymbol\mu) p_{\theta_x}(\mathbf X^{-l}\mid \mathbf z,\boldsymbol\mu)p(\mathbf z,\boldsymbol\mu \mid X_l=x_l,\mathcal D)  d\mathbf X^{-l}d\mathbf zd\boldsymbol\mu,
\end{aligned}
\label{intpos}
\end{align}
where $p(\mathbf z,\boldsymbol\mu \mid X_l=x_l,\mathcal D)$ was approximated by the variational posterior $q(\mathbf z,\boldsymbol\mu \mid X_l=x_l,\mathcal D)$. Since the components in $\mathbf X$ were assume independent, $ p_{\theta_x}(\mathbf X^{-l}\mid \mathbf z,\boldsymbol\mu)$ is derived from $p_{\theta_x}(\mathbf X\mid \mathbf z,\boldsymbol\mu)$ by excluding the $X_l$ component.

% Thus, 
% \begin{align}
% \begin{aligned}
%     P(\mathcal H_0\mid \mathcal D) &= P\big[p^{y_1y_2(x_l)}
%     =p^{y_1y_2(x_{ref})}\big]\\
%     &=\mathbb E\big[ \mathbb I_{\{p^{y_1y_2(x_l)}=p^{y_1y_2(x_{ref})}\}}\big]\approx \frac{1}{n}\sum_{ij} \mathbb I_{\{p^{y_{ij1}y_{ij2}(x_l)}=p^{y_{ij1}y_{ij2}(x_{ref})}\}}.
%     \end{aligned}
% \end{align}
The integrals in Equation \eqref{intpos} are difficult to obtain; hence, we use Monte Carlo samples to estimate the posterior predictive probabilities. We reject $\mathcal H_0$ if the credible interval of the ACE does not contain zero. 

Specifically in this application, for categorical covariates, such as a child's place of residence, maternal education level, the baseline level was chosen as $x_{ref}$ and the other conditions were chosen as $x_{l}$. Moreover, for continuous covariates, such as the child's age, the $x_{ref}$ was chosen as the least observed value.

\section{Simulation study}
We used synthetic data and real data to evaluate the performance of the framework in an ablation study. We generated samples from the hierarchical model shown in Equation \eqref{hierarchical}. 

Specifically, spatially structured high-dimensional data with correlated binary outcomes using a copula-based dependence structure was generated. Let $n$ denote the number of samples. Each individual $i = 1, \dots, n_j$ is associated with spatial location $j=1,2,...,126$. We define a neighborhood graph using the West African spatial map to construct the precision matrix $\mathbf Q$. $d$-dimensional latent variables $\boldsymbol\mu \in \mathbb{R}^d$ that capture spatial dependence. For each latent dimension $k = 1, \dots, d$, we generate:
\[
\boldsymbol\mu_{k} \sim \mathcal{N}(\mathbf 0,(\mathbf D - \rho \mathbf A)^{-1}) \text{ and }  z_{ijk}\sim \mathcal{N}(\mu_{jk},\sigma^2_{jk}),%, \quad \varepsilon^{(k)} \sim \mathcal{N}(0, I_n),
\]
where $\rho = 0.9$ controls the strength of spatial autocorrelation. We introduced dependence across latent dimensions given as $z_{ijk} \leftarrow z_{ijk} + z_{ij(k-1)}, \quad k = 2, \dots, d$. We generate observed features $\mathbf x_{ij} \in \mathbb{R}^p$ from the latent structure $\mathbf x_{ij} = P \mathbf z_{ij} + \boldsymbol\epsilon_{ij}, \quad \boldsymbol\epsilon_{ij} \sim \mathcal{N}(0, \sigma^2 I_p)$, where $P \in \mathbb{R}^{p \times d}$ is a loading matrix with entries sampled from a standard normal distribution.

We consider two binary outcomes $(Y_{i1}, Y_{i2})$. Their marginal probabilities are defined via logistic regression on the latent variables:
\[
\eta_{ij1} = \mathbf z_{ij}^\top \beta_{11}+\sin(\mathbf z_{ij})^\top \beta_{12}, \quad \eta_{i2} = \mathbf z_{ij}^\top \beta_2,
\]
\[
\pi_{ij1} = \frac{1}{1 + e^{-\eta_{ij1}}}, \quad
\pi_{ij2} = \frac{1}{1 + e^{-\eta_{ij2}}},
\]
where $\beta_{11},\beta_{12}, \beta_2 \sim \mathcal  N(\mathbf 0,I_d)$ are coefficient vectors. To model dependence between the two outcomes, we employ a Gumbel copula. Let $(U_{ij1}, U_{ij2}) \sim C_\alpha$, where $C_\alpha$ is a Gumbel copula with parameter $\alpha \geq 1$. The observed binary outcomes are obtained via inverse probability transformation:
\[
Y_{ij1} = \mathbb{I}(U_{ij1} \leq \pi_{ij1}), \quad
Y_{ij2} = \mathbb{I}(U_{ij2} \leq \pi_{ij2}).
\]

This construction ensures that $Y_{ij} \sim \text{Bernoulli}(\pi_{ij})$ marginally, while dependence between $Y_{i1}$ and $Y_{i2}$ is governed by the copula parameter $\alpha$.

\begin{figure}
    \centering
    \includegraphics[scale=0.415]{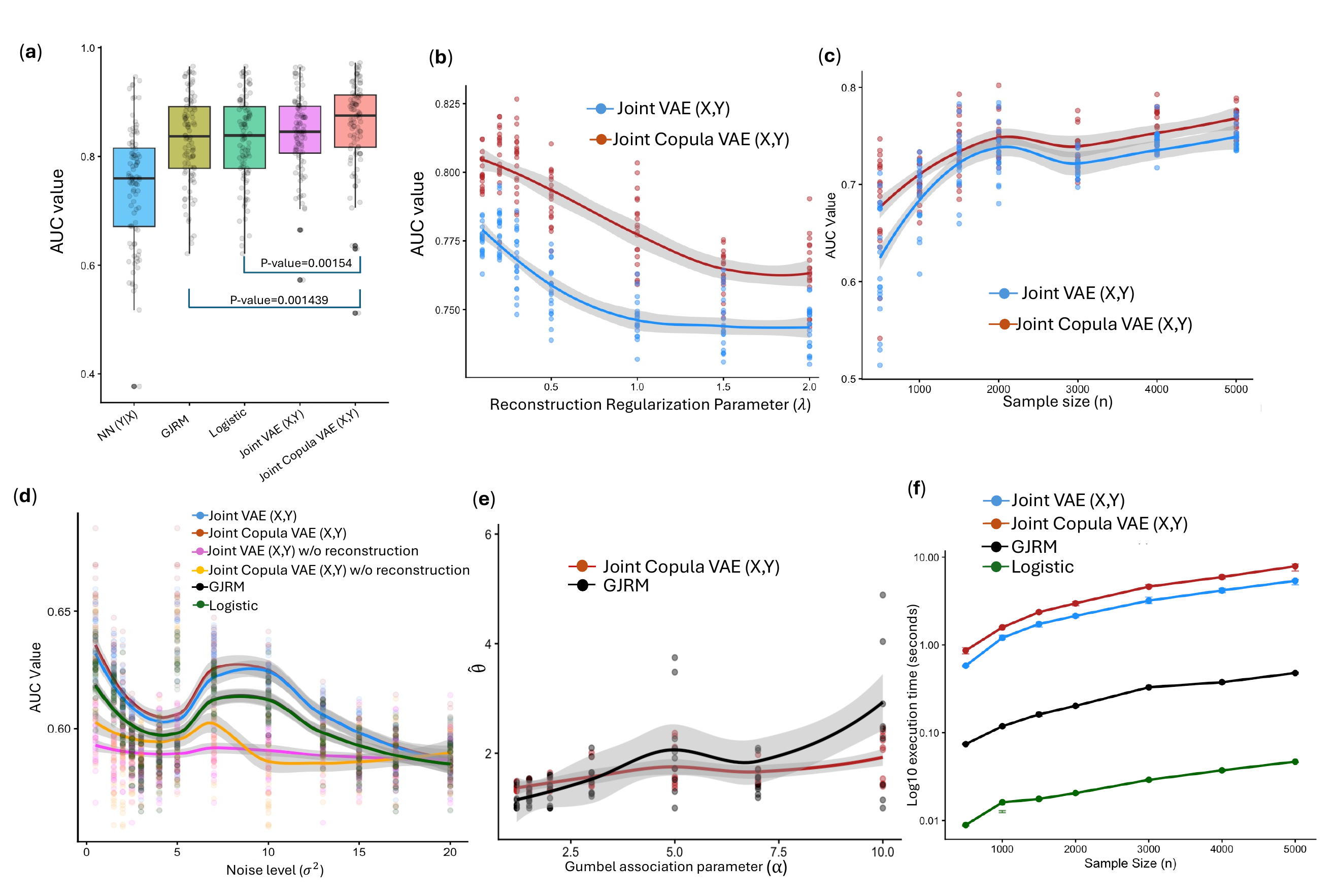}
    \caption{Simulation results. (a) Distribution of the area under the curve (AUC) for the competing models. Higher value indicates better performance. (b) Effect of the tuning parameter $\lambda$ on predictive performance for the joint models. (c) Impact of sample size on model performance. (d) Effect of increasing noise in the covariates on predictive performance. (e) Estimated versus true copula association parameter. (f) The log10 of the execution time in seconds at varying sample sizes.}
    \label{fig:simulationablation}
\end{figure}

We performed an ablation study of the proposed model and benchmarked it against independent logistic regression and Generalized Joint Regression Modeling (GJRM) with Gumbel copula from the \texttt{gjrm} R package \citep{marra2017gjrm}. First, setting $\theta_y=\alpha=2$, $\sigma^2_{jk}=\sigma^2=\lambda=1$, we generated $n=5000$ samples and split into 80\% training and 20\% test sets, with the test set containing both seen and unseen locations. We evaluated the performance of five models, including \texttt{Logistic}, \texttt{GJRM}, \texttt{NN(Y|X)}, \texttt{Joint VAE(Y, X)}, and \texttt{Joint Copula VAE(Y, X)} using the Area under the Curve (AUC). A higher value of AUC indicates a better performance. We used \texttt{Joint Copula VAE(Y|X)} to denote the proposed model, and  \texttt{Joint VAE(Y, X)} to denote the proposed model but treating the $Y_1$ and $Y_2$ as independent (i.e., without a copula). We used \texttt{NN(Y|X)} to denote a feedforward neural network, treating the target responses as independent. In addition, the \texttt{Logistic} model was fitted assuming independence between $Y_1$ and $Y_2$. Figure \ref{fig:simulationablation}a shows the AUC for all the models in the test set, with \texttt{Joint Copula VAE(Y, X)} statistically performing better than the baseline copula model \texttt{GJRM} and independent \texttt{Logistic} model. This VAE's superior performance is likely due to its ability to approximate the non-linear $\sin$ relationship, which the baseline models missed.

%We also investigated how $\lambda$ in Equation \eqref{elbo2} affects the prediction capacity between \texttt{Joint VAE(Y, X)} and \texttt{Joint Copula VAE(Y, X)}, and the result is shown in Figure \ref{fig:simulationablation}b.  The AUC decreases as $\lambda$ increases for both models, but faster for \texttt{Joint VAE(Y, X)}. The AUC levels off at $\lambda=1.5$. While modeling X and Y jointly is beneficial overall, the result shown here implies that modeling the joint could impact the overall prediction performance, especially when the covariate is noisy.

We further investigated the effect of the tuning parameter $\lambda$ in Equation \eqref{elbo2} on predictive performance for both \texttt{Joint VAE(Y, X)} and \texttt{Joint Copula VAE(Y, X)}. $\lambda$ weights the covariate reconstruction loss relative to the predictive likelihood. The results are presented in Figure \ref{fig:simulationablation}b. For a fair comparison, identical stopping criteria were used for both models. Across both approaches, the AUC decreases as $\lambda$ increases, with a more pronounced decline observed for \texttt{Joint VAE(Y, X)}. The performance stabilizes around $\lambda = 1.5$. These findings suggest that while joint modeling of $\mathbf X$ and $\mathbf Y$ is generally beneficial, excessive weighting of the covariate reconstruction term can adversely affect predictive accuracy. This effect is particularly evident when the covariates are noisy, as the model may overemphasize reconstructing $\mathbf X$ at the expense of learning features most relevant for predicting $\mathbf Y$. 

To assess robustness to noise in the covariates $\mathbf X$, we conducted an additional experiment, with results presented in Figure \ref{fig:simulationablation}d. Noise was introduced by perturbing each covariate as $h_{ij} = x_{ij} + \epsilon_{ij}, \quad \epsilon_{ij} \sim \mathcal{N}(0, \sigma^2)$ where the variance was varied over $\sigma^2 \in \{1, 1.5, 2, 2.5, 3, 3.5, 5, 7, 10, 13, 15, 20\}$. Thus, all models were fitted using the noisy covariates $\mathbf H$ instead of the original $\mathbf X$. The results show that the GJRM model and the logistic regression model exhibit nearly identical performance across all noise levels. In contrast, the joint and copula-based models consistently achieve higher AUC values and demonstrate greater robustness to increasing noise. Although their performance exhibits mild oscillatory behavior as noise increases, it eventually stabilizes and converges to that of the competing methods under very high noise levels. Overall, these findings suggest that joint modeling of $\mathbf X$ and $\mathbf Y$ provides improved resilience to covariate noise, likely due to the latent representation acting as a denoising mechanism. 

In additional experiments, we investigated the effect of varying the sample size and the copula dependence parameter, as shown in Figure \ref{fig:simulationablation}c \& e. The results indicate that predictive performance improves with increasing sample size, suggesting desirable statistical consistency (Figure \ref{fig:simulationablation}c). Furthermore, although the proposed model does not precisely recover the true value of the copula parameter, the estimated values exhibit a clear monotonic relationship with the ground truth, and appear similarly with the standard \texttt{GJRM} package (Figure \ref{fig:simulationablation}c). This indicates that the model is able to capture the underlying dependence structure, even if exact parameter recovery is not achieved. The observed discrepancy is likely attributable to the binary nature of the response variables, which limits the information available for accurately estimating dependence parameters. Figure \ref{fig:simulationablation}f presents the $\log_{10}$ of execution time (in seconds) as a function of sample size. As expected, the VAE requires more computation time than the logistic model; however, it remains scalable to very large datasets (including millions of observations) due to its use of stochastic gradient descent, which enables efficient mini-batch optimization.

In addition, we evaluated the behavior of the model to capture the covariate's contribution to the prediction probability. The results are shown in the Supplementary material.  The findings indicate that, although the exact covariate coefficients in classical regression cannot be recovered due to the latent representation inherent in the VAE framework, the estimated ACEs reliably quantify the direction and magnitude of each covariate's influence on the target prevalence. This highlights the model's robustness in capturing covariate epidemiological effects and spatial variability even in high-dimensional settings.  

%While the joint models generally performed better than the conditional and independent models, the y

\section{Results}
 We applied the proposed model to estimate and map the probabilities of comorbidity among children in West Africa, focusing on three pairwise combinations: ``diarrhea and fever", ``ARI and fever", and ``ARI and diarrhea". We focused on pairwise comorbidities because each disease pair may reflect a distinct epidemiological pathway and association. Specifically in the observed data, a three-way outcome is rare and may lead to lower power. The dataset was randomly partitioned into training (80\%) and testing (20\%) subsets. Model optimization was performed using the Adam optimizer, implemented through the \texttt{torch} interface in R~\citep{chowdhry2025deep}, enabling efficient stochastic gradient descent and automatic differentiation. For the encoder and decoder networks, four hidden layers were adopted, each with 60, 30, 20, and 10 units for the encoder and 10, 20, 30, 60 units for the decoder, and 10, 5, 3 units for the predictive network. The estimates of the spatial precision for ``diarrhea and fever", ``ARI and fever", and ``ARI and diarrhea" are $\tau$ = 1.540, $\tau$ = 1.002, and $\tau$ = 0.980. Figure~{\color{red}4} in the supplement illustrates the overlay of comorbidity outcomes on the latent mean space ($\boldsymbol\mu$) learned by the proposed model. The visualization reveals a distinct region of the latent space in which children experiencing comorbid conditions are concentrated, indicating that the model effectively separates comorbid and non-comorbid profiles based on shared underlying risk patterns driven by the socioeconomic, demographic, and spatial conditions.

\subsection{Comorbidity Probability}
\begin{figure}
    \centering
    \includegraphics[scale=0.35]{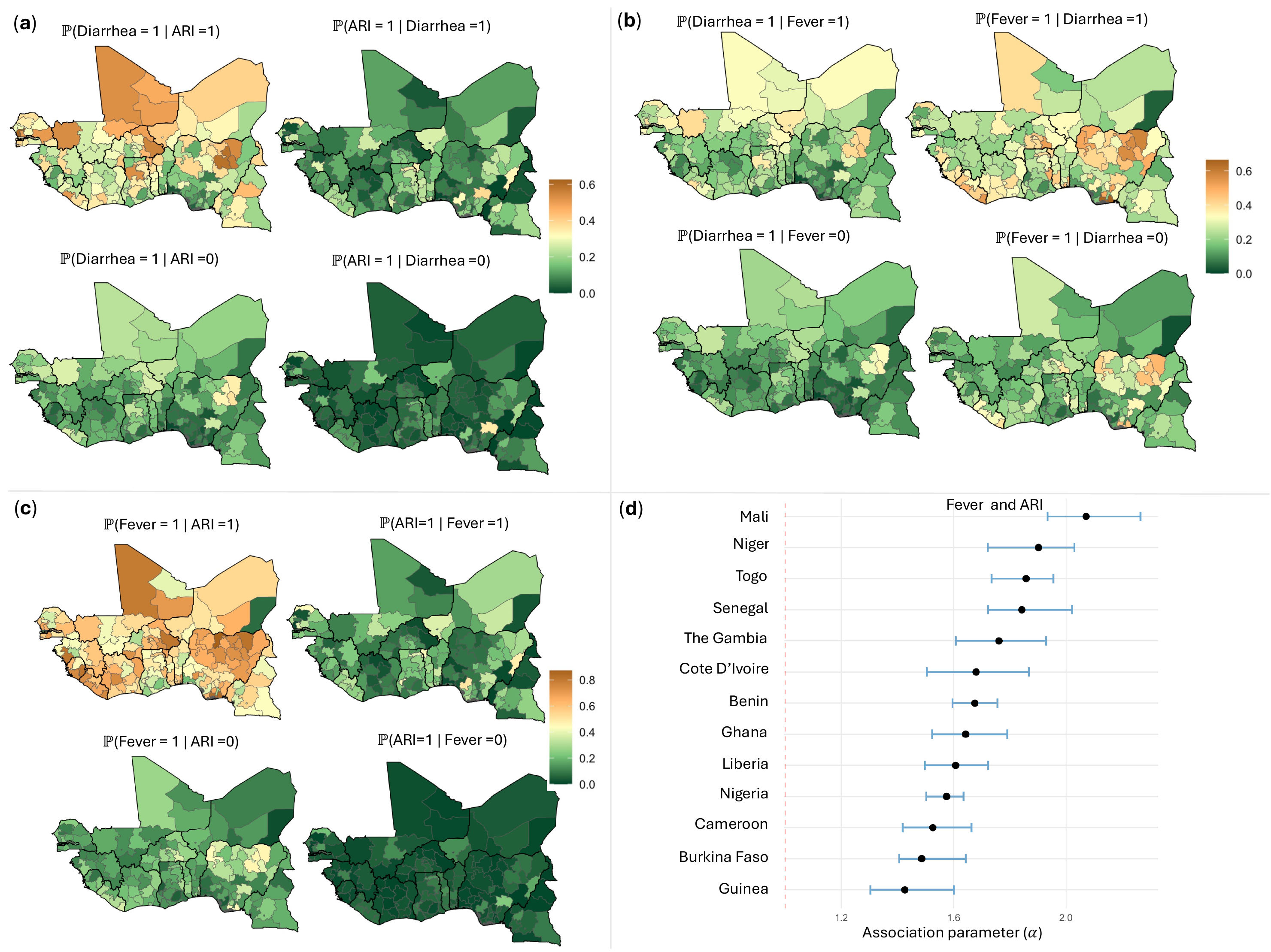}
    \caption{Predicted conditional probabilities of multiple ailments and the country-specific association parameters.}
    \label{fig:cl}
\end{figure}

%[in progress]
Figure 2 in the supplementary material shows the predicted spatial distribution of the joint probability of these health conditions. Figure \ref{fig:cl} presents the corresponding conditional probabilities and estimates of the interaction parameter by country. The global association parameter estimate, $\hat\alpha$, and the 95\% confidence interval for diarrhea and fever are 1.574 (1.549, 1.604), for fever and ARI are 1.691 (1.653, 1.726), and for diarrhea and ARI are 1.357 (1.334, 1.382). This result suggests that the correlation between Fever and ARi is relatively strongest.

Figure \ref{fig:cl}a illustrates the spatial distribution of conditional probabilities for Diarrhea and Acute Respiratory Infection (ARI), highlighting a significant co-occurrence between the two conditions across the study region. The probability of a child having Diarrhea given they have ARI ($\mathbb P(\text{Diarrhea}=1 | \text{ARI}=1)$) is notably higher than the inverse ($\mathbb P(\text{ARI}=1 | \text{Diarrhea}=1)$). While the risk of ARI given Diarrhea generally stays below $0.4$, the risk of Diarrhea given ARI frequently exceeds $0.5$ in many subnational areas.  Comparing the top row (presence of a condition) to the bottom row (absence of a condition) reveals a strong positive association. The probability of Diarrhea is substantially higher when ARI is present (top-left) than when it is absent (bottom-left). This suggests that ARI may be a significant risk factor or marker for concurrent diarrheal disease in these populations.  There is pronounced geographic variation, particularly for $\mathbb P(\text{Diarrhea}=1 | \text{ARI}=1)$. The highest risks ($>0.5$) are concentrated in the northern subnational regions, whereas the southern and central areas show relatively lower conditional probabilities.  The maps for $\mathbb P(\text{Diarrhea}=1 | \text{ARI}=0)$ and $P(\text{ARI}=1 | \text{Diarrhea}=0)$ show that in the absence of a co-occurring infection, the independent probability of either disease is low across most of the map, typically falling below $0.2$.

Figure \ref{fig:cl}b shows the spatial distribution of the conditional probabilities between diarrhea and fever. Comparing the top and bottom rows of the figure, there is a positive relationship, meaning that having a fever is associated with a higher probability of diarrhea, and vice versa. The effect is moderate in most parts of the region compared to the maps in Figure \ref{fig:cl}a. Comparing the top row, the risk of having a fever given diarrhea (top-right) is generally higher than the risk of having diarrhea given a fever (top-left). In the figure, high conditional probabilities are more geographically localized than in Figure \ref{fig:cl}a, with northern and south-east districts in Nigeria, south Sierra Leone, Liberia, and Côte d'Ivoire appearing as distinct pockets rather than expansive regions.

Figure \ref{fig:cl}c shows the geospatial heterogeneity in the conditional probabilities of fever and ARI. Comparing the top row (conditions present) to the bottom row (conditions absent), there is a strong and spatially widespread positive relationship between fever and ARI. Both conditional risks are significantly elevated when the other condition is present. There is a stark difference between the top-row maps. A child with ARI has an extremely high probability of also having a fever ($\mathbb{P}(\text{Fever}=1 | \text{ARI}=1)$), widespread across almost all subnational regions (often $>0.6$). In contrast, a child with only a fever is much less likely to have a concurrent ARI ($\mathbb{P}(\text{ARI}=1 | \text{Fever}=1)$ remains mostly below $0.4$). Fever in the absence of ARI is moderately common, particularly in northeast Nigeria subnational regions, showing a higher probability than ARI without fever. This highlights fever as a very general symptom, whereas ARI is a more distinct condition that almost always manifests with fever.

These results reveal a clear hierarchy in the strength and significance of comorbid relationships. The association between ARI and Fever (Figure \ref{fig:cl}c) is fundamentally the strongest and most widespread across the region. This is followed by the relationship between Diarrhea and ARI (Figure \ref{fig:cl}a), while the relationship between Diarrhea and Fever (Figure \ref{fig:cl}b) is the weakest, showing more localized clusters rather than regional trends. Across all conditions, ARI stands out as the most powerful determinant for other comorbid risks. A child with an ARI diagnosis is very likely to experience fever and significantly more likely to also have diarrhea, positioning ARI prevention and management as a cornerstone for reducing overall childhood illness burden and multi-condition comorbidities in West Africa.

Figure \ref{fig:cl}d presents the correlation parameter of fever, and ARI by country, and Figure 2 in the supplementary material shows the estimates for fever and ARI, and Diarrhea and ARI. These estimates were obtained by fitting country-specific models. From Figure \ref{fig:cl}d, the association between fever and ARI is highest in Mali, followed by Niger and Togo, but lowest in Guinea. These findings show that fever and ARI co-occur more often among children in Mali than in any other country in the region, and least often in Guinea.

\subsection{Covariate effect on the joint comorbidities}
To better understand the risk factors of childhood multimorbidity, we examined the effects of both continuous and categorical covariates on the joint probability of experiencing co-occurring conditions such as diarrhea-fever, diarrhea-ARI, and fever-ARI. This presents the estimated average covariate effects and nonlinear exposure-response functions, highlighting how sociodemographic, household, and individual characteristics influence the likelihood of multiple illnesses occurring simultaneously.

\subsubsection{Average effects of child and mother's age}

\begin{figure}
    \centering
    \includegraphics[scale=0.265]{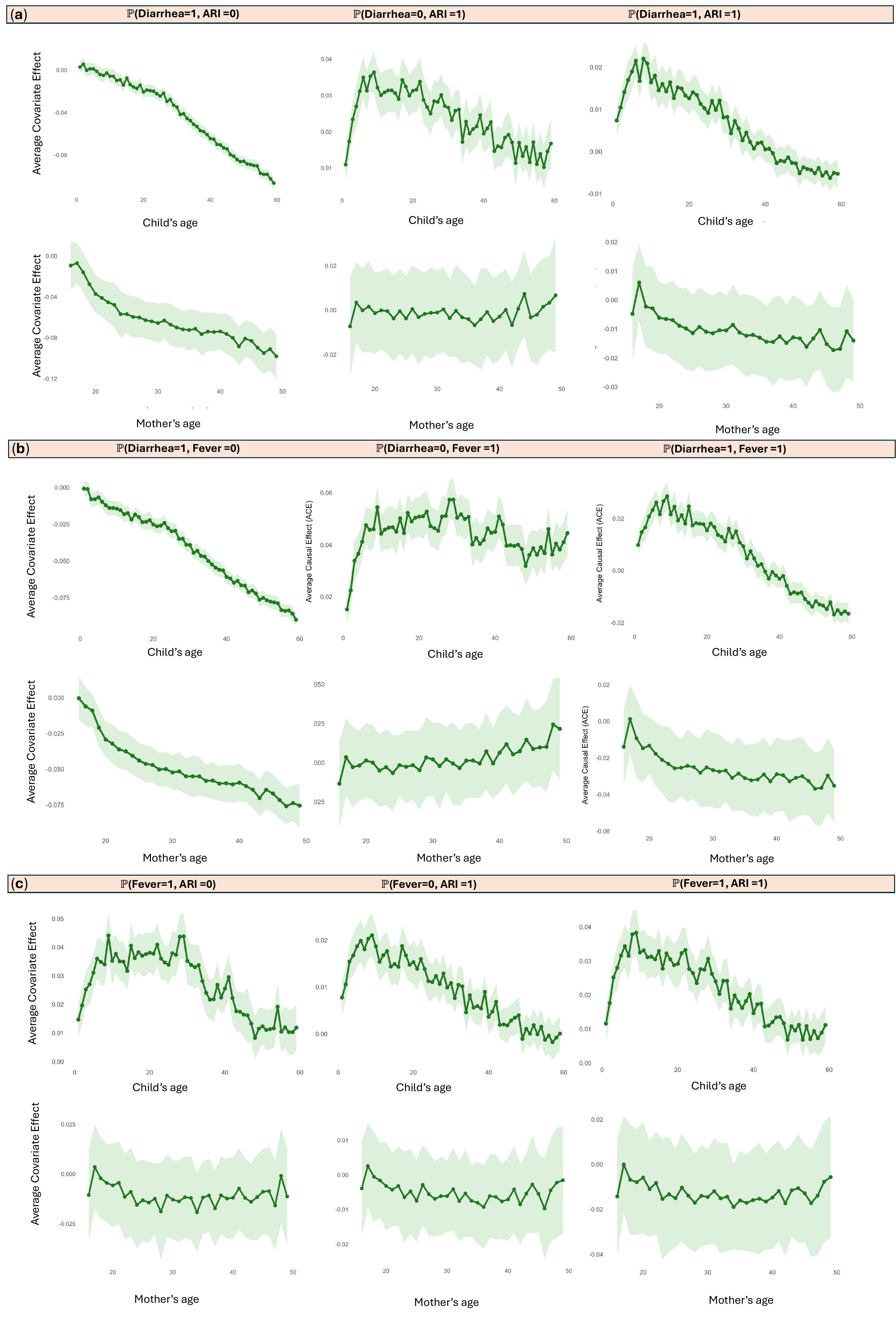}
    \caption{Estimated patterns and the 95\% confidence intervals of the average causal effect across children's and mothers' age on the joint probability of multiple ailments.}
    \label{fig:pl2}
\end{figure}

Figure \ref{fig:pl2} illustrates the nonlinear exposure-condition relationships between maternal and child age and joint childhood morbidities. The Average Covariate Effect (ACE) on the y-axis shows how the likelihood of an outcome shifts as these ages increase.

Child's age shows the most dramatic and consistent patterns across all three panels (Figure \ref{fig:pl2}a-c). For almost all comorbid conditions (the right-most column in each section, where both conditions $=1$), there is a sharp increase in risk that peaks between 10 and 20 months. This likely corresponds to the weaning period and the child's increased exposure to the environment as they become mobile. After the 24-month mark, the risk for all conditions, especially joint infections, steadily declines. By age 5 (60 months), the average effect is at its lowest, suggesting that older children have developed stronger immune defenses. The probability of having only diarrhea ($\mathbb P(\text{Diarrhea}=1, \text{ARI/Fever}=0)$) is highest at birth and drops linearly. This suggests that the youngest infants are most vulnerable to gastrointestinal issues, even without other symptoms. 

In addition, the mother's age generally has a ``protective" effect, though it is less volatile than the child's age. In panels (a) and (b) of Figure \ref{fig:pl2}, as the mother’s age increases, the likelihood of the child having diarrhea generally decreases. This often reflects the benefits of caregiving experience, higher household stability, or better hygiene knowledge that tends to come with age.  In panel (c) of Figure \ref{fig:pl2}, the effect of mother's age on ARI and Fever is relatively flat and stays near zero. The wide confidence intervals suggest that a mother's age isn't as strong a predictor for respiratory issues as it is for diarrheal diseases.
% Maternal age exhibits a modest but systematic nonlinear association across all comorbidity pairs. Elevated risks are observed among children of adolescent mothers (<20 years), while the average effects attenuate toward null as maternal age increases into the mid-20s and early 30s. Beyond approximately 35 years, effects remain close to zero or shift slightly in the protective direction, indicating limited influence at older maternal ages.

% In contrast, child age shows a pronounced and consistent nonlinear effect across all outcomes. Joint morbidity risks are highest during infancy, with sharp declines over the first 12–18 months of life. Thereafter, effects plateau near zero, indicating substantially reduced susceptibility to concurrent diarrhea, fever, and ARI in older children.

% Together, the results reveal early infancy and adolescent motherhood as critical periods of vulnerability for childhood comorbidity, with risk diminishing as maternal maturity increases and child immune development progresses.

% \begin{table}[ht]
% \centering
% \caption{Diarrhea=1 \& ARI = 1}
% \begin{tabular}{rlccc}
%   \hline
%  & var & ACE & lw & up \\ 
%   \hline
% 1 & cough & 0.1060 & 0.1047 & 0.1072 \\ 
%   2 & urban & -0.0018 & -0.0025 & -0.0012 \\ 
%   3 & prim & 0.0047 & 0.0037 & 0.0058 \\ 
%   4 & sec & 0.0010 & 0.0002 & 0.0018 \\ 
%   5 & high & -0.0071 & -0.0084 & -0.0057 \\ 
%   6 & water & -0.0012 & -0.0020 & -0.0003 \\ 
%   7 & toilet & -0.0016 & -0.0024 & -0.0009 \\ 
%   8 & elect & -0.0008 & -0.0015 & -0.0001 \\ 
%   9 & newsp & 0.0009 & -0.0005 & 0.0022 \\ 
%   10 & radio & 0.0025 & 0.0017 & 0.0032 \\ 
%   11 & telv & 0.0018 & 0.0010 & 0.0025 \\ 
%   12 & poorer & -0.0033 & -0.0043 & -0.0023 \\ 
%   13 & middle & -0.0040 & -0.0051 & -0.0029 \\ 
%   14 & richer & -0.0062 & -0.0072 & -0.0051 \\ 
%   15 & richest & -0.0067 & -0.0079 & -0.0055 \\ 
%   16 & work & 0.0019 & 0.0011 & 0.0026 \\ 
%   17 & bord\_23 & -0.0011 & -0.0022 & 0.0000 \\ 
%   18 & bord\_4 & -0.0011 & -0.0020 & -0.0001 \\ 
%   19 & female & -0.0002 & -0.0010 & 0.0005 \\ 
%   20 & breast & 0.0038 & 0.0032 & 0.0045 \\ 
%    \hline
% \end{tabular}
% \end{table}

\subsubsection{Average effects for categorical covariates}
Table \ref{tab:ACEFixed} provides the Average Covariate Effect (ACE) showing how much the probability of a joint condition changes compared to a reference group. If the value is negative, it is a protective factor; if positive, it is a risk factor.

The result shows that socioeconomic factors are the strongest predictors of multimobility. As households move from ``Poorer" to ``Richest," the risk for all joint conditions drops significantly. The Richest households see the largest reduction in risk, particularly for the Diarrhea + Fever combination (ACE = $-0.0187$). While ``Primary" education shows a slight positive risk (perhaps due to confounding factors), Higher Education acts as a major shield, significantly reducing the risk of all three comorbidities. Children in Urban areas have a consistently lower risk of joint illness compared to rural children, likely due to better access to healthcare and infrastructure. Improved water source, improved toilet Type, and access to electricity have negative ACE values across the board. This confirms that basic sanitation and household utility improvements are directly correlated with lower rates of respiratory and diarrheal co-occurrence. Interestingly, children of working mothers show a positive ACE (increased risk). This might suggest a ``supervision gap" or challenges in maintaining consistent feeding and hygiene routines while working. The table shows a slight positive ACE for breastfeeding. In many epidemiological studies, this is an age-confounded result. Breastfeeding is often continued because a child is in the high-risk weaning age (6–24 months) or because the child is currently ill and unable to eat solids. Radio access unexpectedly correlates with a slight increase in risk, while Newspaper and Television generally trend toward lower risk (though some intervals, like TV for Fever+ARI, cross zero and are not statistically significant). Lower-order birth positions (2–3) were modestly protective for some outcomes, whereas higher birth order showed small and inconsistent effects. Child sex exhibited no meaningful association with joint morbidity risk.

% Table \ref{tab:ACEFixed} summarises the average causal effects of categorical covariates on joint childhood morbidities. Overall, socio-economic advantage and improved living conditions were consistently protective across comorbidity pairs, while indicators of exposure and caregiving constraints increased risk.

% Urban residence, higher maternal education, improved water and sanitation, household electricity, mass media exposure via newspapers or television, and increasing household wealth were associated with significantly lower joint probabilities of diarrhea–fever, diarrhea–ARI, and fever–ARI. Protective effects generally strengthened along socio-economic gradients, with the largest reductions observed among children from the wealthiest households and those whose mothers attained higher education.

% In contrast, maternal employment, breastfeeding status, and radio exposure were associated with increased risks for at least two comorbidity pairs, suggesting potential pathways linked to caregiving constraints, exposure patterns, or residual confounding. Lower-order birth positions (2–3) were modestly protective for some outcomes, whereas higher birth order showed small and inconsistent effects. Child sex exhibited no meaningful association with joint morbidity risk.

% Collectively, these results reveal the central role of structural and environmental determinants in shaping childhood comorbidity patterns, with consistent evidence that improvements in household resources, sanitation, and maternal education substantially reduce the burden of multiple concurrent illnesses.

\begin{table}
\centering
\caption{Categorical covariates' effect on the joint comorbidities $(\times 10^{-1})$.} %{\color{red}please multiply all values by 10 and make it three decimal places.}
\resizebox{0.85\textwidth}{!}{%
\begin{tabular}{lccc|ccc|ccc}
\toprule
\textbf{Variable} &
\multicolumn{3}{c}{\textbf{$ (\text{Diarrhea}=1, \text{Fever}=1)$}} &
\multicolumn{3}{c}{\textbf{$ (\text{Diarrhea}=1, \text{ARI}=1)$}} &
\multicolumn{3}{c}{\textbf{$ (\text{Fever}=1, \text{ARI}=1)$}} \\
\cmidrule(lr){2-4} \cmidrule(lr){5-7} \cmidrule(lr){8-10}
 & \textbf{ACE} & \textbf{2.5\%} & \textbf{97.5\%} 
 & \textbf{ACE} & \textbf{2.5\%} & \textbf{97.5\%} 
 & \textbf{ACE} & \textbf{2.5\%} & \textbf{97.5\%} \\
\midrule
 %Cough: \texttt{Yes} & 0.1324 & 0.1307 & 0.1341 & 0.1060 & 0.1047 & 0.1072 & 0.2070 & 0.2050 & 0.2089 \\
 Residence: \texttt{Urban} & -0.078 & -0.088 & -0.068 & -0.018 & -0.025 & -0.012 & -0.053 & -0.068 & -0.039 \\
  Education: \texttt{Primary} & 0.069 & 0.056 & 0.082 & 0.047 & 0.037 & 0.058 & 0.081 & 0.062 & 0.101 \\
  Education: \texttt{Secondary} & -0.027 & -0.037 & -0.016 & 0.010 & 0.002 & 0.018 & 0.019 & 0.004 & 0.033 \\
 Education: \texttt{Higher} & -0.223 & -0.238 & -0.207 & -0.071 & -0.084 & -0.057 & -0.089 & -0.116 & -0.061 \\
 Water source: \texttt{Improved} & -0.044 & -0.055 & -0.033 & -0.012 & -0.020 & -0.003 & -0.027 & -0.042 & -0.012\\
  Toilet type: \texttt{Improved} & -0.045 & -0.055 & -0.036 & -0.016 & -0.024 & -0.009 & -0.044 & -0.058 & -0.031 \\
  Electricity: \texttt{Yes} & -0.073 & -0.083 & -0.064 & -0.008 & -0.015 & -0.001 & -0.029 & -0.042 & -0.016\\
 Newspaper: \texttt{Yes} & -0.044 & -0.063 & -0.0265 & 0.009 & -0.005 & 0.022 & 0.023 & -0.002 & 0.047\\
  Radio: \texttt{Yes} & 0.010 & 0.000 & 0.020 & 0.025 & 0.017 & 0.032 & 0.051 & 0.039 & 0.063 \\
  Television: \texttt{Yes} & -0.025 & -0.035 & -0.015 & 0.018 & 0.010 & 0.025 & -0.000 & -0.013 & 0.013 \\
  Wealth index: \texttt{Poorer} & -0.042 & -0.057 & -0.027 & -0.033 & -0.043 & -0.023 & -0.048 & -0.068 & -0.028 \\
  Wealth index: \texttt{Middle} & -0.074 & -0.087 & -0.062 & -0.040 & -0.051 & -0.029 & -0.058 & -0.076 & -0.041\\
  Wealth index: \texttt{Richer} & -0.123 & -0.135 & -0.110 & -0.062 & -0.072 & -0.051 & -0.094 & -0.113 & -0.075\\
  Wealth index: \texttt{Richest} & -0.187 & -0.201 & -0.173 & -0.067 & -0.079 & -0.055 & -0.104 & -0.122 & -0.086\\
  Maternal work: \texttt{Yes}  & 0.059 & 0.051 & 0.067 & 0.019 & 0.011 & 0.026 & 0.090 & 0.079 & 0.102\\
 Birth order: \texttt{2-3} & -0.026 & -0.038 & -0.014 & -0.011 & -0.022 & 0.000 & -0.006 & -0.022 & 0.011\\
  Birth order: \texttt{4 \& above}& 0.013 & 0.000 & 0.025 & -0.011 & -0.020 & -0.001 & 0.011 & -0.006 & 0.028 \\
  Sex: \texttt{Female} & -0.007 & -0.015 & 0.002 & -0.002 & -0.010 & 0.005 & 0.009 & -0.002 & 0.020\\
  Breast fed: \texttt{Yes} & 0.041 & 0.032 & 0.049 & 0.038 & 0.032 & 0.045 & 0.005 & -0.008 & 0.018\\
\hline
\end{tabular}}
\label{tab:ACEFixed}
\end{table}

\section{Discussion}

This study proposed a novel spatially regularized variational autoencoder framework combined with a Gumbel bivariate copula to jointly model childhood diarrhea, fever, and acute respiratory infection (ARI) across West Africa. Through the integration of spatial dependence, nonlinear covariate effects, and flexible dependence structures, the framework moves beyond conventional single-outcome analyses and enables simultaneous estimation of joint and conditional disease risks. The framework provides a predictive and inference capability and quantification of uncertainty. Simulation studies demonstrated that the proposed approach outperforms conventional methods, while the empirical application revealed substantial spatial heterogeneity among childhood illnesses in the region.

The spatial distribution of marginal disease probabilities reveals a pronounced north–south gradient across West Africa. Diarrhea is more prevalent in northern and Sahelian countries, whereas fever and ARI are more common in southern and coastal regions. This contrast is not immediately apparent from national prevalence statistics alone but becomes evident when spatial patterns are examined at sub-national geographic resolution. The higher burden of diarrhea in northern areas likely reflects complex health and infrastructural limitations and arid climatic conditions that %chronic water scarcity, reliance on unimproved water sources, limited sanitation infrastructure, and arid climatic conditions that 
exacerbate fecal-oral transmission pathways \citep{Kazembe2009,Orunmoluyi2022,Bolarinwa2021,Yaya2018,Nwokoro2020}. In contrast, the elevated prevalence of fever and ARI in southern regions is consistent with humid environments, higher population density, intense malaria transmission, and ecological conditions that favor respiratory and vector-borne infections.

%Joint probability maps further reveal that diarrhea–fever is the most prevalent comorbidity across the fourteen countries studied, with particularly high probabilities concentrated in northern Nigeria, notably in Yobe, Bauchi, and Gombe. The clustering of diarrhea–fever comorbidity in these areas suggests the convergence of multiple vulnerabilities, including poor water and sanitation, low maternal education, food insecurity, and limited access to healthcare services \citep{Kazembe2009,Orunmoluyi2022,Bolarinwa2021,Amadu2023,Yaya2018,Nwokoro2020,Victor2025}. In contrast, diarrhea–ARI and fever–ARI comorbidities are relatively rare across most of the region. Evidence from national and multi-country DHS studies indicates that while diarrhea and ARI are each common among under-five children, their simultaneous occurrence is less frequent, with only a small proportion of children experiencing both conditions concurrently \citep{Mulatya2020,Tekeba2025}. This pattern likely reflects differences in exposure pathways and seasonality, as enteric and respiratory infections often peak under distinct environmental conditions, even though shared vulnerabilities such as malnutrition can increase susceptibility to both \citep{Black2013}.

An important but less immediately visible finding is the tendency for elevated comorbidity risk in districts located near national borders. These border areas often experience weaker health system coverage, high population mobility, informal cross-border trade, and reduced access to consistent water, sanitation, and immunization services. Such structural conditions can amplify disease transmission and complicate surveillance and intervention efforts, leading to spatial clustering of multiple childhood illnesses \citep{Orunmoluyi2022}. This highlights the need for cross-border and regional public health strategies rather than purely national interventions.

The conditional probability results reveal important asymmetries in disease co-occurrence that are not evident from marginal or joint patterns alone. Across countries, diarrhea is substantially more likely among children who already have ARI than ARI is among children with diarrhea. This asymmetry suggests that respiratory infections may weaken immune function or nutritional status, thereby increasing susceptibility to enteric infections. DHS-based studies have similarly shown that although concurrent diarrhea–ARI episodes are less common than single conditions, they remain epidemiologically significant and are strongly associated with shared vulnerability factors, including undernutrition and poor living conditions \citep{Mulatya2020,Tekeba2025}. Spatially, regions with higher overall disease burden tend to exhibit elevated conditional risks, reinforcing the role of local environmental and household contexts in shaping comorbidity patterns \citep{Orunmoluyi2022}.

The strongest conditional dependence is observed between fever and ARI, with fever prevalence consistently high among children with respiratory infections. This aligns with clinical expectations, as fever is a hallmark symptom of pneumonia and other lower respiratory tract infections. However, the reverse relationship, ARI among children presenting with fever, remains weak across most countries, highlighting that fever alone is a poor predictor of respiratory illness. In low-resource settings, limited access to diagnostic tools and laboratory confirmation can exacerbate misclassification and delay appropriate care \citep{Graham2008,Ssentongo2023}.

%Fever shows a stronger conditional association with diarrhea than the reverse, indicating that children with diarrhea are more likely to present with fever than children with fever are to have diarrhea. Fever prevalence is particularly high in malaria-endemic and high-transmission settings across sub-Saharan Africa, and studies using DHS data have documented frequent co-occurrence of fever and diarrhea, with marked subnational variation \citep{Kazembe2007,Ssentongo2023}. %Multi-country analyses further indicate that respiratory infections and diarrhea are among the strongest correlates of febrile illness in under-five children, with respiratory illness typically showing the strongest association, followed by diarrhea \citep{Ssentongo2023}. Underlying factors such as malnutrition can further compromise immune defenses and increase the likelihood of multiple infections \citep{Carvalho2025,Black2013}.
These findings reveal the limitations of symptom-based diagnosis in endemic settings, where fever is common but non-specific and may obscure underlying comorbidities, leading to inappropriate or delayed treatment \citep{Graham2008,Ssentongo2023}.

The estimated average covariate effects (ACEs) provide further insight into the associations or drivers of spatial heterogeneity in comorbidity risk. Younger maternal age, particularly below 20 years, is associated with modestly elevated risks for all comorbidity combinations. This likely reflects a combination of limited health knowledge, socioeconomic disadvantage, and reduced access to maternal and child health services among adolescent mothers \citep{frempong_2022,mutuku_2020}. ACEs decline toward zero among mothers aged 25–34, a group typically characterized by improved caregiving capacity, better nutritional status, and greater engagement with preventive healthcare services \citep{berhan_2025}. Among older mothers, ACEs approach zero or become slightly negative, suggesting a small protective effect potentially driven by accumulated caregiving experience and greater household stability \citep{ezra_2023}.

Child age exhibits the strongest nonlinear effect, with infants under one year facing the highest comorbidity risk. Immature immune systems, exposure during weaning, and unsafe feeding practices likely contribute to this vulnerability \citep{diah_2025}. The sharp decline in ACEs during the first 12–18 months corresponds to immune maturation and the protective effects of routine immunization against major ARI pathogens, such as \textit{Haemophilus influenzae} type B and \textit{Streptococcus pneumoniae} \citep{larbi_2025}. By toddlerhood, comorbidity risk stabilizes at low levels, reflecting improved adaptive immunity and hygiene practices \citep{ezra_2023}.

Socioeconomic and environmental covariates further explain observed spatial patterns. Higher maternal education, improved water and sanitation, electricity access, and household wealth consistently reduce comorbidity risk, reinforcing evidence that structural living conditions are central to childhood health outcomes \citep{frempong_2022,niels-hugo_2020,sanni_2018,fischer_2011,sharlene_2021}. The similarity in spatial trajectories between northern Nigeria and parts of Liberia, despite geographic separation, suggests that shared structural vulnerabilities, rather than proximity, drive comorbidity risk. Both settings are characterized by poverty, fragile health systems, and limited access to safe water and sanitation, illustrating how common constraints can generate parallel epidemiological patterns across distant regions \citep{Amadu2023,Victor2025}.

Despite the promise of the proposed model, it has some limitations. While ACEs offer a population-level summary of covariate influence, they represent marginal effects over the latent manifold rather than the conditional log-odds common in frequentist GLMs. In addition, the proposed model requires substantial retraining and cross-validation for every new dataset and new problem, which could limit its accessibility. Though the framework has the potential to make predictions at fine spatial resolution, the current implementation does not directly make predictions at such resolution. Future research will focus on refining the architecture to enhance automated scalability and enable direct fine-scale spatial inference.

%Thus, these findings demonstrate that childhood illnesses in West Africa are deeply interconnected, spatially clustered, and shaped by a combination of environmental, socioeconomic, and health-system factors. By jointly modeling disease dependence and spatial heterogeneity, this study reveals patterns that are not readily observable from marginal analyses alone. The results reveal the importance of integrated intervention strategies that address shared determinants—particularly in high-burden regions such as northern Nigeria and border districts—and highlight the need for coordinated, cross-border public health responses.

\section{Conclusion}
This study introduces a novel spatially regularized variational autoencoder framework integrated with a Gumbel copula to model the joint distribution of diarrhea, fever, and acute respiratory infection among children in West Africa. Through jointly capturing the complex spatial structure and asymmetric dependence between health outcomes, the proposed approach provides a flexible and interpretable method for mapping childhood comorbidity risk. Our findings reveal substantial subnational heterogeneity in childhood comorbidity risk across West Africa, revealing the need for geographically targeted interventions rather than uniform national strategies. The produced maps can guide health ministries and development partners in prioritizing high-burden regions for intensified child health programs. The strong conditional dependence observed, particularly fever given diarrhea or ARI, highlights the importance of integrated management of childhood illnesses within primary healthcare, including routine screening for co-occurring symptoms to improve early detection and treatment. Additionally, the elevated risk among infants indicates that early-life interventions, such as postnatal care, growth monitoring, vaccination, and caregiver education, should remain central to child health policy. Finally, mixed effects of media exposure suggest the need to refine health communication strategies, optimizing content and delivery across platforms to ensure accurate and effective messaging reaches caregivers in high-risk settings.
% \section*{Acknowledgments}
% The authors acknowledge the Demographic and Health Survey program for granting access to the data used in this study.

% \subsection*{Author contributions}
% OAE conceived the study idea, developed the model, and performed benchmarking analysis. EG sourced the data, and OAE, BDI, and FE performed the real data analysis and interpreted the results. All authors participated in the manuscript writing, and EG revised it critically for intellectual content. All authors read and accepted the final manuscript.\\
% \subsection*{Financial disclosure}

% No funding was received.

% \subsection*{Conflict of interest}

% The authors declare no potential conflict of interest.
% \subsection*{ Data accessibility}
% The data used in this study are publicly available on the Demographic and Health Survey Program website at \href{https://dhsprogram.com/}{https://dhsprogram.com/}.

%\nocite{*}% Show all bib entries - both cited and uncited; comment this line to view only cited bib entries;

\bibliography{References}%

\clearpage

%\section*{Author Biography}

% \begin{biography}{\includegraphics[width=66pt,height=86pt,draft]{empty}}{\textbf{Author Name.} This is sample author biography text this is sample author biography text this is sample author biography text this is sample author biography text this is sample author biography text this is sample author biography text this is sample author biography text this is sample author biography text this is sample author biography text this is sample author biography text this is sample author biography text this is sample author biography text this is sample author biography text this is sample author biography text this is sample author biography text this is sample author biography text this is sample author biography text this is sample author biography text this is sample author biography text this is sample author biography text this is sample author biography text.}
% \end{biography}

\end{document}